# Field – Induced Spin Phase Transitions in the Cobalt Valence Tautomers

A.K.Zvezdin[a], A.S.Mischenko[b,*]

[a]Institute of General Physics of RAS, 119991, Moscow, Russia
[b]KOFEF, Physics Department, M.V.Lomonosov Moscow State University, 119899, Moscow, Russia

**Abstract.** Magnetic properties of the *Co* valence tautomers are considered in wide range of magnetic field and temperature. Particularly, a model of the first order structural phase transition accompanied by a magnetic moment jump is proposed. The phase diagrams of the phase transition on a *B-T* plane (*B* is magnetic field and *T* is temperature) are calculated for four ligand types. Dependence of the critical magnetic field on physical characteristics known from experiments (transition temperature at the absense of magnetic field, entropy change per formula unit, energy difference between two phases) is found. The phase diagram is parametrized for comparison with other phase transitions of this kind.

This work is supported by RFBR grants № 02-02-17389, № 99-02-17830, INTAS (99-01839).

**1. Introduction.**

Field induced phase transitions (FIPT) attract great attention as they represent a useful tool for direct studying of the basic interactions which are responsible for materials' magnetic structure and for definition of the parameters of these interactions [1-6]. The research direction in this area has been changing from studying bulk materisals to studying nanostructures and high-spin molecules (mesoscopic molecules like *$Mn_{12}Ac$, $Fe_8$, $Fe_{10}$*, nanoclusters, carbon nanotubes, etc.) recently [7-11]. Compounds experiencing so-called spin phase transition (*low spin - high spin*) are quite interesting both from fundamental and applied points of view. Among them are the materials with the phase transition of several types: $\gamma\text{–}Ce^{3.06+}$ - $\alpha\text{-}Ce^{3.67+}$, $Co^{III}$-$Co^{II}$, $Mo^{III}$-$Mo^{II}$, the phase transition in $Fe^{2+}$ with spin change from *S = 0* to *S = 2*, etc [12-23]. These compounds are promising for nanotechnology in general and for the

---

[*] Corresponding author: **smischenko@yahoo.com**

*2D* - systems (thin films and surfaces containing nanoclusters) and nanocomposite materials with nanoclusters inclusions in particular.

This work is dedicated to study the cobalt valence tautomers. Tautomerism is a coexistence of several molecular isomeric forms (states). Such tautomeric molecules can be found in different states depending on some peculiar conditions but the whole system is in dynamic equilibrium. These molecules are called tautomers and such a transition is called tautomeric (or the spin transition in our case) [24]. The energy of two isomeric states vs. the distance between ligands is plotted on fig.1 in order to illustrate the nature of the tautomeric phase transition.

The tautomeric molecules have unstable electronic structure due to proximity of two tautomeric forms' energy levels, so the molecules are quite sensitive to external perturbations. Photons, electric field, magnetic field, pressure can change the electronic state of the tautomeric molecules. The polycristal $Fe^{II}$ samples experience the light-induced low spin – high spin phase transition at low temperatures ($T$ < 50 K). Magnetization hysteresis on temperature has been observed at room temperatures in [27-29]. The described tautomeric molecular complexes are nanoclusters with linear size about 10-100 nm.

The considered spin phase transition (which is the first order phase transition as usual) is accompanied by the change of the magnetic atoms' valence, substantial change (sometimes as large as 20%) of the interatomic distance inside each tautomeric molecule, change of the sample color and sometimes magnetization hysteresis on temperature in the vicinity of the transition. As a rule, there are the following «initiators» of the phase transition: temperature change, pressure change and light irradiation. But it can be induced by magnetic and electric fields as well. Study of the phase transition in strong magnetic fields shows us these processes from a different point of view. Besides, we can find out the nature of the phase transition more thoroughly and determine the parameters which are responsible for it. Applications of the molecular complexes are quite diverse. First, they can be used as elementary memory devices in computers due to the magnetization hysteresis on temperature [23]. Second, the color change accompanying the phase transition can be applied in the visualization systems. For instance, the multi-coloured indicator has been created out of the materials with *«Fe»* spin transition [27-29].

We are studying the *B-T* phase diagram of the transition, the magnetization vs. *T* and the *B* dependence for molecules with different ligands and compare the properties of the spin phase transition in the *Co* valence tautomers with that of other compounds studied before (*Ce* and *YbInCu$_4$*). The phase diagrams of the latter compounds have been studied theoretically in [22] and experimentally in [32-37]. The phase transition in *Ce* (so-called $\alpha$-$\gamma$ phase transition) is the closest to the tautomeric transition from the point of view of basic physical characteristics and phase diagram curve. In our case the critical magnetic field is 100-200 T which is about the same in *Ce*.

It is necessary to point out that recently there has been a substantial progress in the *FIPT* study of the bulk materials and nanoclusters in the megagauss fields range [38-47].

**2. Physical and chemical properties of the *Co* valence tautomers**

The general chamical formula of a *Co* valence tautomer can be represented as *[Co(3,5-DTBSQ)$_2$(N^N)]*, where *3,5-DTBSQ* stands for semiquinonate form 3,5-di-tert-butyl-o-quinone, the following compounds can be used by way of a *N^N* ligand: *phen* (1, 10 - phenanthroline); *bpy* (2, 2' - bipyridine); *dmbpy* (4, 4' - dimethyl-2,2'-bipyridine); *dpbpy* (4, 4' - diphenyl - 2, 2' - bipyridine). All the four types of tautomers exhibit a low-spin *[Co$^{III}$(3,5-DTBSQ).)CAT-(N^N)]* - high-spin *[Co$^{II}$(3,5-DTBSQ)$_2$ (N^N)]* phase transition, where *(3,5-DTBCAT)$^{2-}$* is a catecholate form of an o-quinone - ligand [16]. The high spin phase is a mixed valent state. From the physical point of view this means that the molecular ground state in this phase is not pure itself but is a superposition of several pure quantum states with different valencies. The detailed description of the electronic structure is given in the next paragraph.

The *Co$^{III}$* valence tautomer synthesis technique is described in [16-21]. The *Co$^{III}$ (low spin)* - *Co$^{II}$ (high spin)* phase transition induced by temperature change has the following features. First, there is a jump of magnetic moment of each molecule. Second, this phase transition is accompanied by the substantial change of the *Co-ligands, ligand-ligand* and *Co-Co* distances (it reaches 10% or 2 $\overset{o}{A}$). The *Co* valence changes as well as the net negative electric charge in the vicinity of a *Co* atom (negative electric charge is transferred from the *Co* ion on ligands) what is confirmed by the susceptibility and optical absorbtion measurements [14]. This is an entropy driven process and the entropy change reaches values of 1-2 *R* per mole according to the heat capacity measurements [13]. The phase transition temperature $T_{1/2}$ varies broadly depending on ligand type ($T_{1/2\ MIN} = 226.6\ K$, $T_{1/2\ MAX} = 350.0\ K$, see table 1) [16].

| Ligand type | $T_{1/2}$, K | $\Delta E$, K | $B^0_{Crit}$, T |
|---|---|---|---|
| Phen | 226.6 | 189.7 | 141.2 |
| Bpy | 277.0 | 231.9 | 172.5 |
| Dmbpy | 286.6 | 239.9 | 178.5 |
| Dpbpy | 350.0 | 293.0 | 218.0 |

**Table 1.** Transitions temperatures $T_{1/2}$ [14], energy difference between the ground states of the two phases $\Delta E$ (calculation according to (12) for $\Delta S_{vib} = 1.2\ R$) and critical values of magnetic field $B^0_{Crit}$ (see formula (13)) for different ligand types.

**3. Phase diagram of the field induced low spin - high spin transition in the Co valence tautomers**

The electron structure of a *Co* ion with ligands is split in two groups - $t_{2g}$ and $e_g$, (see fig.2) which results from the *Co* 3d-orbital energy splitting. The $t_{2g}$ levels are lower than $e_g$ so we can say that the $t_{2g}$ levels are shielded from the surroundings. The quantum states belonging to the $e_g$ energy group mix with the ligand $\sigma$-orbitals thus forming molecular orbitals of different kinds and configurations. The energy levels split leads to stabilization of high-spin configuration at high enough temperatures. In these conditions there are two electrons on the $e_g$ level and one hole on $t_{2g}$. The situation is quite different at low temperatures: the $e_g$ level is unoccupied and the $t_{2g}$ level is fully occupied, one electron from the *Co* 3d-orbital is transferred onto the surrounding ligands. Thus, the net electric charge is larger in low-temperature phase.

The interactions hierarchy of the *Co* valence tautomers is in the following (in descending order): (a) electrostatic interaction between valency *Co* 3d-electrons, (b) spin-orbital interaction between these electrons, (c) electrostatic interaction between *Co* and ligand electrons. The last interaction results in electron jump from a *Co* atom to the nearest ligand (and conversely) at the phase transition. The first two interactions are responsible for the formation of the net magnetic moment of a *Co* ion for the fixed number of electrons on its valency *d*-orbital.

A typical feature of the considered phase transition is a changing number of electrons on a *Co* ion and the nearest ligands. The following formalism is usually used to describe the composite system "*Co ion + ligands*":

$$\hat{H} = \hat{H}_A + \hat{H}_d + \hat{H}_z \qquad (1)$$

where $\hat{H}_A$ is an Anderson impurity model hamiltonian [14, 48], $\hat{H}_d$ stands for a hamiltonian describing spin-spin interactions between *Co 3d*-orbital electrons and $\hat{H}_z$ describes the interaction with the external magntic field.

$$\hat{H}_A = -\sum_{i\alpha,j\beta}\sum_{\sigma} t_{i\alpha,j\beta}\hat{c}^+_{i\alpha\sigma}\hat{c}_{j\beta\sigma} + U_d \sum_{\gamma\sigma \neq \gamma'\sigma'} \hat{n}_{\gamma\sigma}\hat{n}_{\gamma'\sigma'} \qquad (2)$$

$$\hat{H}_d = J_H \sum_{\substack{\gamma>\gamma' \\ \sigma>\sigma'}} \hat{\vec{s}}_{\gamma\sigma}\hat{\vec{s}}_{\gamma'\sigma'} + \sum_{i=1}^{N_d} \xi(\vec{r}_i)\hat{\vec{l}}_i\hat{\vec{s}}_i \qquad (3)$$

$$\hat{H}_z = -\mu_B \vec{B} \sum_{i\alpha} \left(\hat{\vec{l}}_i + 2\hat{\vec{s}}_i\right) \qquad (4)$$

where $t_{i\alpha,j\beta}$ are hopping integrals for $i\alpha \neq j\beta$ and on-site energies for $i\alpha = j\beta$; $i$ and $j$ are site indices, which include one *Co* ion, two nitrogen atoms belonging to the *N-N* complex and four oxygen atoms; $\alpha$, $\beta$ are orbital symmetry-adapted labels; $c^+_{i\alpha\sigma}$ ($c_{j\beta\sigma}$) is a creation (annihilation) operator of an electron on site $i$ ($j$) with orbital moment $\alpha$ ($\beta$) and spin projection $\sigma$, $U_d$ is the direct Coulomb integral and $J_H < 0$ is the ferromagnetic Hund exchange integral; $n_{\gamma\sigma} = d^+_{\gamma\sigma}d_{\gamma\sigma}$ is the occupancy operator in the *Co 3d*-orbital $\gamma$ with spin projection $\sigma$, $\xi(r_i)$ is the free-ion parameter. Several of the required parameters of the model hamiltonian described above can be found in literature and the rest of them can be calculated. The on-site electron energies and approximate values of the hopping integrals are listed in the reference [49]. Hoping matrix elements can be represented as linear combinations of the above hopping integrals according to the Slater-Koster theory [50]. We would like to point out that the hamiltonian (1) describes both electron-electron interactions on a *Co* ion for $i = j$ and hybridization of ligand *s*-electron orbital and *Co d*-electron orbital for $i \neq j$. Exactly the last interaction is responsible for mixed valence states of the system. Another approach to description of such phase transitions concernes spin-phonon coupling. This method is physically supported by a number of reasons: first, crystal lattice parameters change crucially at the phase transition and second, this phase transition can be induced by external pressure [22].

The *Co* ions are surrounded by ligand crystal electrostatic field of cubic symmetry, what leads to splitting of multiplets belonging to *3d $^7$- (3d $^6$-) Co* configurations in high spin (low spin) phase. This is the reason of partial degeneracy raising of the ground state and nearest excited states. The significant contibution to high spin phase is made by terms *$^4T_1$* of the configuration *$3d^7$ ($S_{Co}$ = 3/2, $S_L$ = 0, $S_{Net}$ = 3/2)* and *$^5T_2$* of the configuration *$3d^6$ ($S_{Co}$ = 2, $S_L$ = 1/2, $S_{Net}$ = 3/2)*, and to low spin phase - by the only one term *$^1A_1$* of configuration *$3d^6$ ($S_{Co}$ = 0, $S_L$ = 1/2, $S_{Net}$ = 1/2)*, where $S_{Net}$ is the net spin of the whole molecule (fig. 2). The described model allows to explain the enthalpy *ΔH ~ 0.03 eV/molecule* jump which corresponds to the entropy jump *ΔS ~ 1-2 R* calculated from heat capacity measurements [13].

We are using these configurations and their wave functions to calculate the eigenvalues of the hamiltonian describing interaction of the system with external magnetic field.

In order to investigate the influence of external magnetic field on the processes taking place in the system in the vicinity of the phase transition, the entropy of low spin and high spin phases can be presented as

$$S^{(\alpha)} = S^{(\alpha)}_{magn} + S^{(\alpha)}_{vib}, \qquad (5)$$

where $S^{(\alpha)}_{magn}$ stands for magnetic part of entropy and $S^{(\alpha)}_{vib}$ is a vibrational part of the total entropy. The latter covers such processes as electron jump from a *Co* ion to ligands, change of the *Co-Co* and *Co-ligand* distances (intramolecular processes) and change of the molecules' vibrational spectra in a crystalline matrix (intermolecular processes). The index $\alpha$ takes two values - $\alpha$ = *Low spin* and $\alpha$ = *High spin*.

The magnetic part of the entropy ( *S = (E - F)/T, F = - kT ln Z* ) is

$$S^{(\alpha)}_{magn}(B,T) = \frac{\varepsilon^{(\alpha)}}{T} + k \ln Z^{(\alpha)}_{magn}(B,T) \qquad (6)$$

where $\varepsilon^{(\alpha)}$ is the intrinsic energy per molecular unit depending on intermolecular parameters (the net molecular spin, the distance between ligands in molecules and between molecules themselves, magnetic ion valency, etc.) and independent of temperature and magnetic field; $Z^{(\alpha)}_{magn}(B,T)$ stands for the partial function of a molecule, describing its interaction with the magnetic field.

The free energy per one molecular unit depends on entropy like

$$F^{(\alpha)} = \varepsilon^{(\alpha)} - TS^{(\alpha)} = -kT \ln Z_{magn}^{(\alpha)}(B,T) - TS_{vib} \qquad (7)$$

The partial function for both phases can be calculated from the Zeeman spectra of a free ion:

$$E^{LS} = E_{LS} + g\mu_B \widetilde{S}_z^{LS} B_z,$$
$$E^{HS} = E_{HS} + g\mu_B \widetilde{S}_z^{HS} B_z, \qquad (8)$$

where $E^{LS}$ ($E^{HS}$) is the energy of the high spin (low spin) phase taking into account the Zeeman splitting, $E_{LS}$ ($E_{HS}$) are the energies independent from magnetic field, $\widetilde{S}^{LS} = 1/2$ ($\widetilde{S}^{HS} = 3/2$) are spins of the phases, $B$ stands for external magnetic field, $g = 2$ is the hyrotropic ratio. Thus, the free energy of each phase can be represented as

$$F_{LS} = -kT \ln\left(2e^{-\frac{E_{LS}}{kT}} ch\frac{\mu_B B}{kT}\right) - TS_{vib}^{LS}$$

$$F_{HS} = -kT \ln\left(e^{-\frac{E_{HS}}{kT}} \frac{sh\frac{4\mu_B B}{kT}}{sh\frac{\mu_B B}{kT}}\right) - TS_{vib}^{HS} \qquad (9)$$

The phase transition line is determined by the following equation:

$$F_{LS}(B,T) = F_{HS}(B,T) \qquad (10)$$

After substituting (9) into (10) and obvious transformation we get the formula for the phase transition line in a B-T plane:

$$B = \frac{1}{2}\frac{kT}{\mu_B} Arch\left(\frac{1}{2}\exp\left(\frac{\Delta E - T\Delta S_{vib}}{k_B T}\right)\right), \qquad (11)$$

where $\Delta E = E_{HS} - E_{LS}$, $\Delta S_{vib} = S_{vib}^{HS} - S_{vib}^{LS}$ are intrinsic energy and entropy jumps at the phase transition. Besides, we can obtain some useful information out of (11) by substituting zero values for magnetic field ($B=0$) and temperature ($T \to 0$), respectively:

$$\frac{\Delta E}{T_{1/2}} - \Delta S_{vib} = k \ln 2 \qquad (12)$$

$$B^0{}_{Crit} = \frac{1}{2}\frac{\Delta E}{\mu_B} \tag{13}$$

where $T_{1/2}$ is the transition temperature at the absense of the magnetic field and $B^0{}_{Crit}$ is the transition magnetic field value at ultra-low temperatures.

We would like to point out that $\Delta S_{vib}$ can depend on temperature and this dependence will not influence on the outward appearance of the obtained formulas. But because of ambiguous experimental results concerning the entropy value and its behaviour we take it constant on temperature, namely $\Delta S_{vib}$ = 1.2 R per mole [14].

The next step of this work is parametrization of the formula (11) for comparison of the considered phase transition with *Ce* phase transition described in [12]. For this purpose we use dimensionless ($\tau, \beta$) and parametrization ($\varphi$) variables: $\tau = T/T_{1/2}$, $\beta = B/B^0{}_{Crit}$, $\beta/\tau = tg\varphi$  Then by means of well known trigonometric formulas we obtain:

$$\beta^2 + \tau^2 = \frac{\tau^2}{\cos^2\varphi} \tag{14}$$

Expression (14) in terms of the introduced variables takes the form:

$$\beta^2 + \tau^2 = R(\varphi) \tag{15}$$

where

$$R(\varphi) = \left[\frac{\dfrac{\Delta E}{k_B T_{1/2} \cos\varphi}}{\ln\left\{2ch\left(\dfrac{2\mu_B B^0{}_{Crit}}{k_B T_{1/2}} tg\varphi\right)\right\} + \dfrac{\Delta S_{vib}}{k_B}}\right]^2 \tag{16}$$

The constants $\Delta E/k_B T_{1/2}$ and $\mu_B B^0{}_{Crit}/k_B T_{1/2}$ are the same for all the ligands, as follows from formulas (12) and (13), so the function $R(\varphi)$ is the same for all the ligands as well (see fig.4). The deviation of the function $R(\varphi)$ from a constant value does not exceed 10%, what is 4% higher than in case of *Ce* phase transition [12]. In other words, the phase diagram in dimensionless variables can be considered as a circle $\beta^2 + \tau^2 = const$ with 10% precision. The difference between *Co* and *Ce* cases can originate from difference in vibrational modes of the crystalline system taken into account in this work.

There are different values for $\Delta E$ and $\Delta S_{vib}$ in literature and some of them are conflicting (see, for instance, [13, 14, 16]). However, this may be due to different determination of $T_{1/2}$ from experimental data.

The magnetization *M* of the system can be calculated according to $M(B,T) = -\partial F^{(\alpha)}(B,T)/\partial B$, where $F^{(\alpha)}(B,T)$ is the free energy of the equilibrium phase (9).

$$M_{LS} = \mu_B \tanh\left(\frac{\mu_B B}{k_B T}\right)$$

$$M_{HS} = \mu_B \tanh\left(\frac{\mu_B B}{k_B T}\right) \frac{5 + \tanh^2\left(\frac{\mu_B B}{k_B T}\right)}{1 + \tanh^2\left(\frac{\mu_B B}{k_B T}\right)} \quad (17)$$

Magnetization vs. magnetic field dependence at T = 150 K for two types of ligands - *"Phen"* and *"Dpbpy"* is plotted on fig. 5. It is obvious that magnetization of compounds with different ligands differ from each other only in the magnetic field value at which the magnetization jump occurs. That is the reason why there are only two curves plotted on this figure.

The above calculations are aproximate and have pretensions of only qualitative correspondence with future experiments. Using more precise values of $\Delta E$ and $\Delta S_{vib}(T)$ will substantially improve the calculation results. Fundamentally, our model can be improved in the following directions. First, it is necessary to consider structure of each phase more thoroughly. Then it is interesting to perform a detailed calculation of electron jump from *Co* ion to ligands in magnetic field accompanied by change of the ion valency.

**4. Conclusion**

The *B-T* phase diagram of the *Co* valence tautomers is obtained. The critical values of the magnetic field at ultra low temperatures are within the range from 140 T to 220 T depending on the ligand type. The phase diagram is shown to have the same functional dependence as in the *Ce* at the *α-γ* phase transition or in the *"Kondo"* material *YbInCu$_4$*.

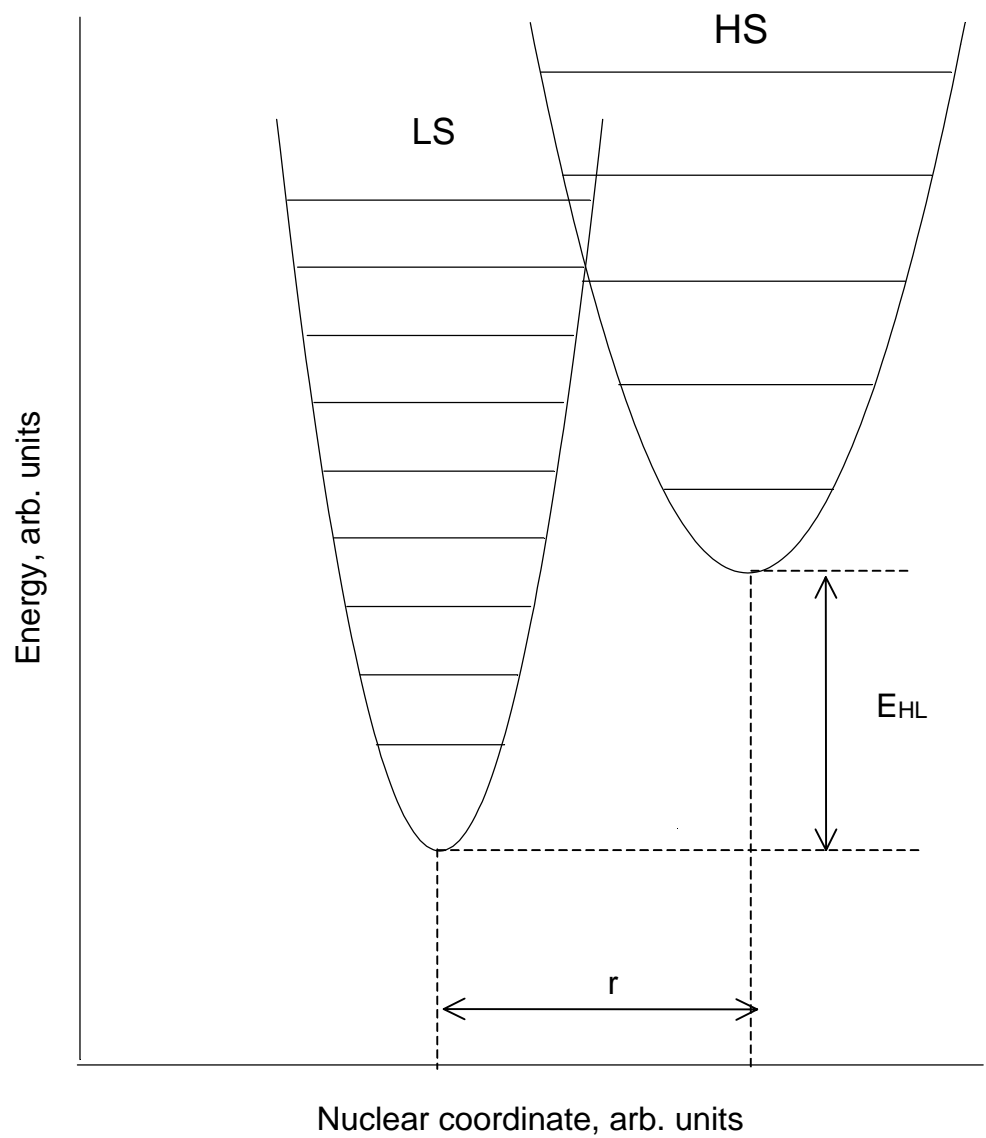

**Fig. 1.** The high spin and low spin phase energies vs. the *Co*-ligand distance.

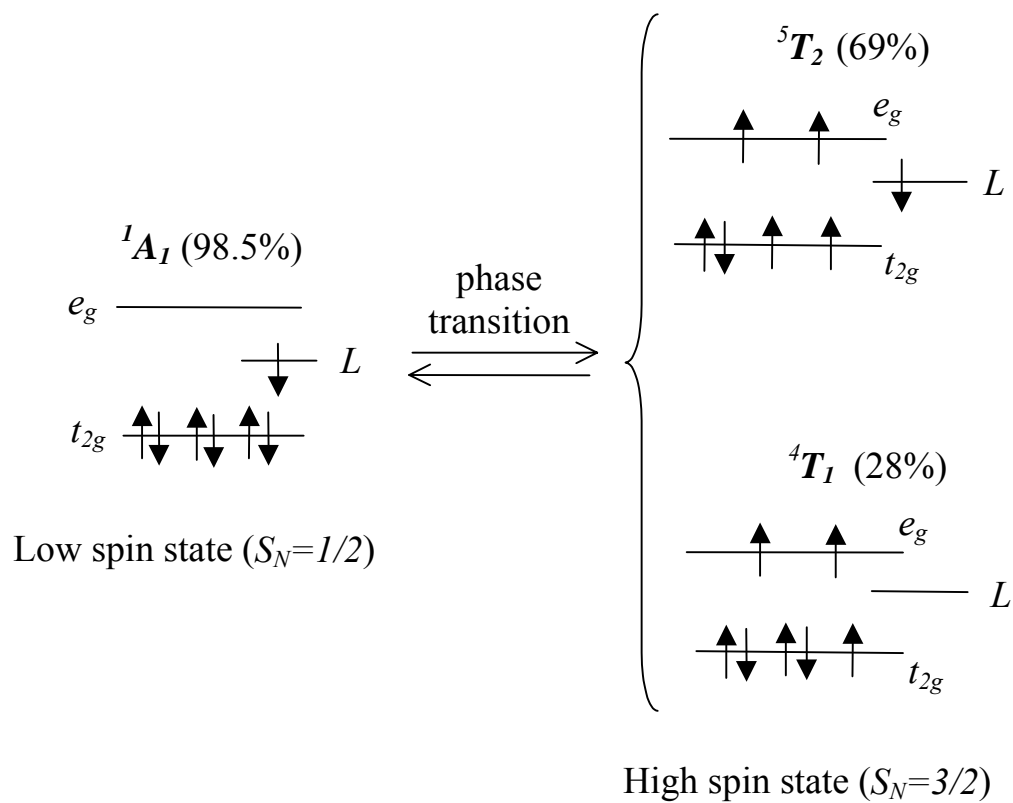

**Fig. 2.** The $Co^{2+}$ ion *d*-electrons' distribution over the energy levels in the quantum system *"$Co^{3+}$ + ligand"*. This scheme illustrates the competition between the Hund interaction in a *d*-ion which tends to create the state with maximum spin of the system and interaction with the ligands which tends to destroy the Hund ferromagnetic state. The high spin state is a mixed state - it mainly consists of two terms $^5T_2$ and $^4T_1$ with the weights 0.69 and 0.28 respectively.

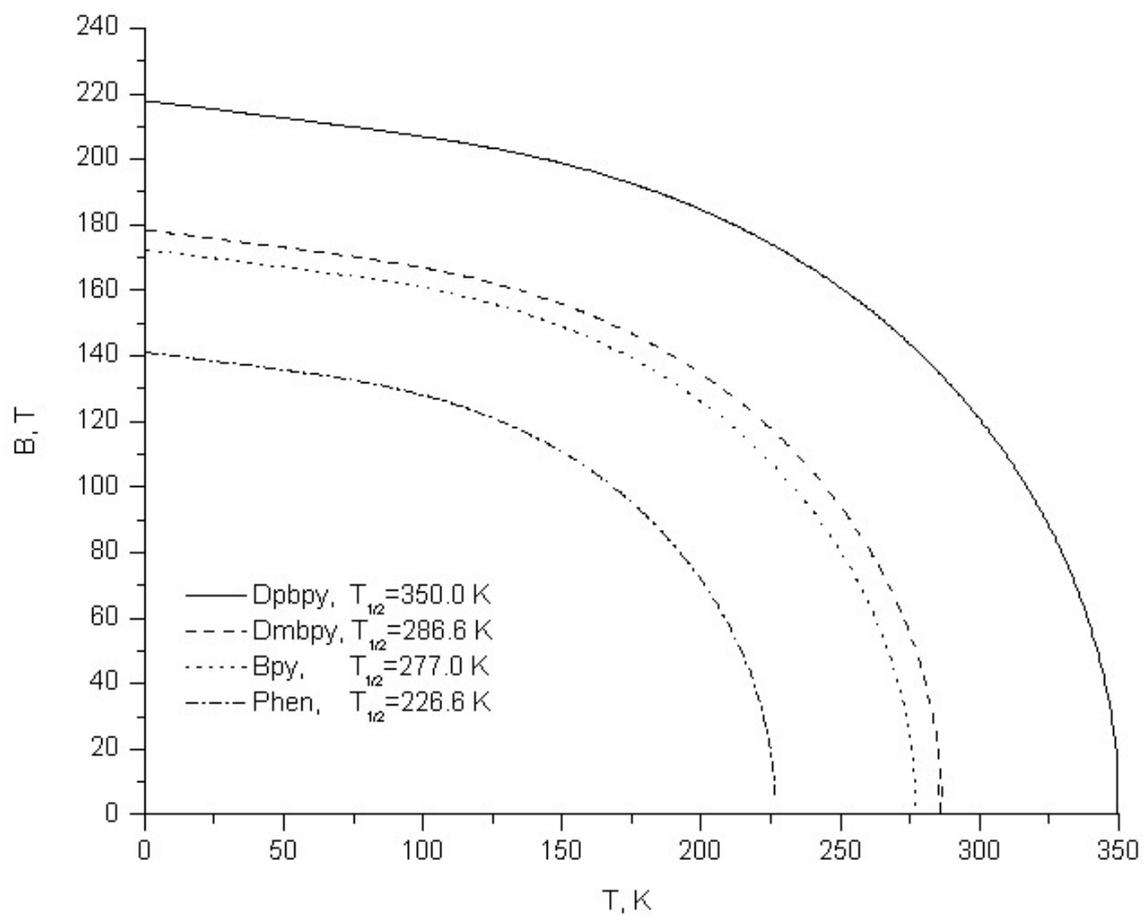

**Fig. 3.** The *B-T* phase diagram of the considered spin phase transition in *Co* valence tautomers.

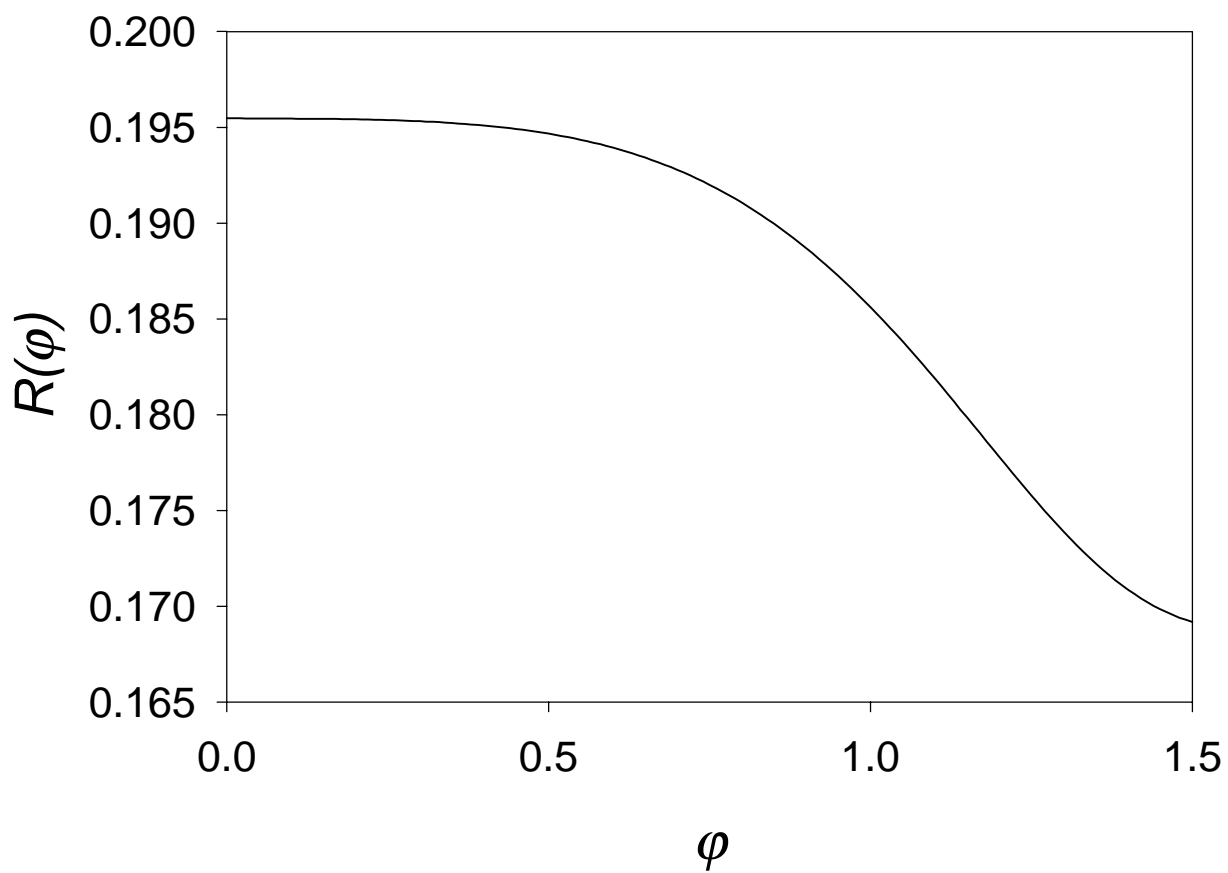

**Fig. 4.** *R(φ)* function, calculated for the valence tautomers (see formula (16) in the text). This function is a constant with 10% precision only.

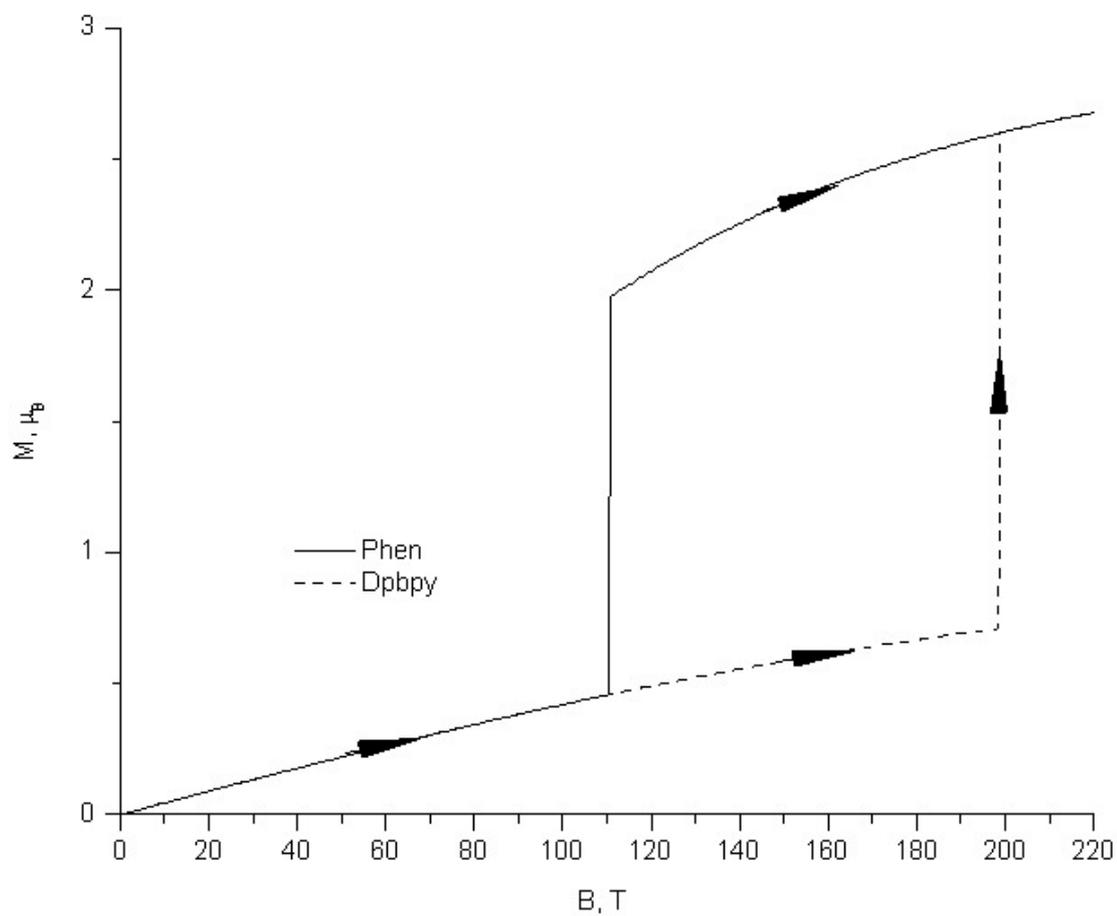

**Fig. 5.** Magnetization of two tautomers with the ligands "Phen" and "Dpbpy" vs. temperature in the vicinity of the phase transition at T = 150 K.